\documentclass[twocolumn]{aastex7}
\usepackage{amsmath}
\usepackage{placeins}
\graphicspath{{figures/}}
\shorttitle{Binary Microlensing in 3C 279}
\shortauthors{Yang et al.}
\providecommand{\restartappendixnumbering}{}

\begin{document}

\title{A Candidate Binary Microlensing Event in the 2018 Gamma-Ray Flare of 3C 279}

\author{Yihong Yang} 
\affiliation{Shanghai Astronomical Observatory, Chinese Academy of Sciences}
\affiliation{School of Astronomy and Space Sciences, University of Chinese Academy of Sciences} 
\email[show]{yangyh@shao.ac.cn} 

\author{Jianeng Zhou} 
\affiliation{Shanghai Astronomical Observatory, Chinese Academy of Sciences}
\email[show]{zjn@shao.ac.cn}

\author{Haibin Ren}
\affiliation{Department of Astronomy, Tsinghua University, Beijing 100084, China}
\email{rhb23@mails.tsinghua.edu.cn}

\author{Wei Zhu}
\affiliation{Department of Astronomy, Tsinghua University, Beijing 100084, China}
\email{weizhu@mail.tsinghua.edu.cn}

\author{Haiyun Zhang}
\affiliation{Department of Physics, Key Laboratory of Astroparticle Physics of Yunnan Province, Yunnan University, Kunming 650091, People's Republic of China}
\email{zhanghy@ynu.edu.cn}

\author{Dahai Yan}
\affiliation{Department of Physics, Key Laboratory of Astroparticle Physics of Yunnan Province, Yunnan University, Kunming 650091, People's Republic of China}
\email{yandahai@ynu.edu.cn}

\author{Hongping Deng}
\affiliation{Shanghai Astronomical Observatory, Chinese Academy of Sciences}
\email{hpdeng353@shao.ac.cn}

\author{Hengxiao Guo}
\affiliation{Shanghai Astronomical Observatory, Chinese Academy of Sciences}
\email{hxguo@shao.ac.cn}

\author{Zhen Yan}
\affiliation{Shanghai Astronomical Observatory, Chinese Academy of Sciences}
\email{zyan@shao.ac.cn}

\author{Liang Chen}
\affiliation{Shanghai Astronomical Observatory, Chinese Academy of Sciences}
\email{chenliang@shao.ac.cn}

\author{Minfeng Gu}
\affiliation{Shanghai Astronomical Observatory, Chinese Academy of Sciences}
\email{gumf@shao.ac.cn}

\begin{abstract}
The April 2018 gamma-ray flare of the flat-spectrum radio quasar 3C~279 exhibited sharp peaks within a broader outburst. Its rapid substructure was previously attributed to magnetic reconnection. Motivated by its distinctive temporal profile within our comparison sample of bright blazar flares and its resemblance to caustic-crossing features in optical microlensing events, we investigate a finite-source binary-microlensing interpretation. A jointly fitted model of binary-lens magnification, a smooth intrinsic flare, and Gaussian-process variability reproduces the principal double-peaked structure of the GeV light curve. The adopted local solution has a binary mass ratio of $q\simeq0.098$ and an Einstein timescale of $t_{\rm E}\simeq0.327$~d. The absolute lens mass and projected source radius depend on the lens distance and relative proper motion. The fitted geometry can correspond to a Galactic planetary-mass binary or an intervening stellar or substellar system, with different physical source sizes in each case. We present binary microlensing as a testable interpretation of this event, with the finite-source structure offering a potential probe of the compact GeV-emitting region.
\end{abstract}

\section{Introduction}\label{sec:introduction}
Rapid gamma-ray variability in blazars probes compact emitting regions within relativistic jets, but their angular sizes remain far below the resolution of gamma-ray telescopes \citep{Ackermann2016}. Gravitational microlensing offers a complementary probe: near a caustic, the magnification depends on the source size and surface-brightness distribution \citep{Paczynski1986,Wambsganss2006,Torres2003}. In the strongly lensed blazar B0218+357, the gamma-ray flux ratio between the lensed images differs from the radio ratio and varies with time. These properties have been interpreted as additional microlensing by stars in the lensing galaxy, yielding constraints on the size of the gamma-ray emitting region \citep{VovkNeronov2016}. For a binary lens, the caustic geometry allows multiple magnification peaks along a single source trajectory \citep{Wambsganss1997}.

The flat spectrum radio quasar 3C 279, at redshift $z=0.536$, underwent a bright gamma-ray outburst in April 2018. Its ``peak-in-peak'' structure and rapid subflares were interpreted as emission from plasmoids formed by magnetic reconnection \citep{Shukla2020}. The sharp peaks and distinctive temporal profile of this event also motivate a binary-microlensing interpretation, in which a foreground binary lens modulates a smooth intrinsic flare with correlated stochastic variability.

A microlensing origin would change the physical interpretation of the rapid variability: the sharp structure would reflect the lens geometry and source motion rather than equally rapid intrinsic energy release. It would also connect the flare profile to the angular size of the GeV emitting region. We examine whether a finite-source binary-lens model can reproduce the observed profile and whether the inferred lens and source configurations are physically plausible.

\section{LAT Flare Morphology}\label{sec:data}

We analyzed \textit{Fermi} Large Area Telescope (LAT) data over modified Julian date (MJD) 58219--58233 in the 0.1--300 GeV band. A binned likelihood analysis with \texttt{Fermitools} v2.2.0 and \texttt{Fermipy} \citep{Wood2017} provided the time-resolved flux and photon-index estimates. Event selections, the region model, and the alternative time bins are described in Appendix~\ref{app:lat}. The light curve binned into 2500 s was used for the lens fit.

\begin{figure}[t]
\centering
\includegraphics[width=\columnwidth]{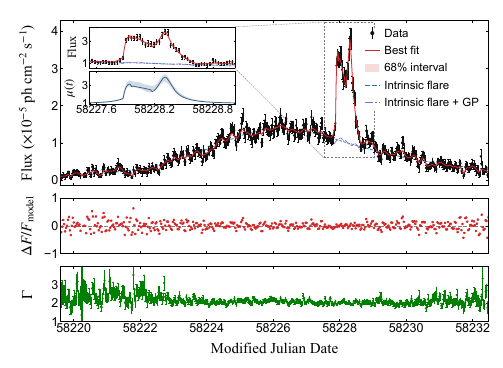}
\caption{\textit{Fermi}-LAT light curve and the best-fit joint GP--lens model. \textbf{Top:} The 0.1--300 GeV flux measurements in 2500 s bins (black), the underlying intrinsic flare (blue dashed line), the intrinsic flare plus GP component (purple dash-dotted line), and the best-fit joint model (red solid line), with its 68\% pointwise interval (red shading). The inset enlarges the boxed interval, showing the light curve (upper) and magnification with its 68\% interval (lower) over the trajectory range shown in Figure~\ref{fig:geometry}. \textbf{Middle:} Relative flux residuals, $(F_{\rm obs}-F_{\rm model})/F_{\rm model}$. \textbf{Bottom:} The photon index $\Gamma$ obtained from the power-law model.}
\label{fig:fit}
\end{figure}

The sharp peak within the broader outburst comprises two distinct sub-peaks, separated by an interval on the order of a few hours (Figure~\ref{fig:fit}). The structure remains visible with several different time binnings. During MJD 58227.67319--58228.62213, the photon index shows no clear variation and remains consistent with that measured over the broader flare ( Figure~\ref{fig:flare-score-index}(b); Appendix~\ref{app:spectral-index}).

We also compare the event with 57 selected LAT flares: 54 from other blazars and three other flares of 3C 279. For a smooth double-exponential fit (Equation~\ref{eq:flare}) to each selected window, define the normalized peak amplitude $A=\max(F_{\rm fit})/\langle F_{\rm obs}\rangle$ and the empirical score $S=A/(\tau_{\rm rise}+\tau_{\rm decay})$, with the timescales in days. The 2018 event has $S=5.84$~d$^{-1}$, markedly higher than the values in the comparison sample, most of which lie below $3$~d$^{-1}$ (Figure~\ref{fig:flare-score-index}(a); Appendix~\ref{app:comparison}).

\section{Binary-lens Model and Fit}\label{sec:model}

We represent the mean intrinsic emission as a constant plus a smooth double-exponential flare \citep{Abdo2010},
\begin{equation}
F_{\rm int}(t)=F_{\rm c}+F_0\left[
\exp\!\left(-\frac{t-t_{\rm p}}{\tau_{\rm rise}}\right)+
\exp\!\left(\frac{t-t_{\rm p}}{\tau_{\rm decay}}\right)
\right]^{-1}.
\label{eq:flare}
\end{equation}
Here $t_{\rm p}$ is a reference epoch, not generally the maximum of an asymmetric flare, and $F_0$ sets its normalization. The observed flux is
\begin{equation}
F_{\rm obs}(t)=\mu(t)\,[F_{\rm int}(t)+g(t)]+\eta(t),
\label{eq:observed}
\end{equation}
where $\mu(t)$ is the time-dependent magnification factor introduced by the microlensing. The zero-mean Gaussian process $g(t)$ describes correlated intrinsic variability with a stochastically driven, damped simple-harmonic-oscillator (SHO) kernel \citep{ForemanMackey2017}, and $\eta(t)$ represents measurement noise. The smooth flare and GP share the same magnification, assuming that both arise in the same magnified region.

The binary-lens model is described by seven parameters (Table~\ref{tab:lensparams}): the closest-approach epoch $t_0$, the impact parameter $u_0$, the Einstein timescale $t_{\rm E}$, the source radius $\rho$, the projected lens separation $s$, the mass ratio $q=M_2/M_1\leq1$, and the angle $\alpha$ between the source trajectory and the binary axis. The angular distances corresponding to $u_0$, $\rho$, and $s$ are normalized by the Einstein radius $\theta_{\rm E}$ of the total lens mass, with $\rho=\theta_\star/\theta_{\rm E}$. The timescale is defined as $t_{\rm E}=\theta_{\rm E}/\mu_{\rm rel}$, where $\mu_{\rm rel}$ is the relative angular speed of the source and lens. Including the five intrinsic-flare quantities and three SHO hyperparameters ($S_0$, $Q$, and $\omega_0$) gives 15 fitted parameters. We approximate the projected gamma-ray emitting region as a circular disk of uniform surface brightness.
We compute the finite-source binary-lens magnification using \texttt{VBBinaryLensing} \citep{Bozza2010,Bozza2018}.
The model parameters are optimized using CMA-ES, followed by Markov chain Monte Carlo sampling to characterize the local posterior distribution. Details of the parameter-space search, optimization, and sampling are given in Appendix~\ref{app:lenssearch}.

\begin{figure}[t]
\centering
\includegraphics[width=\columnwidth]{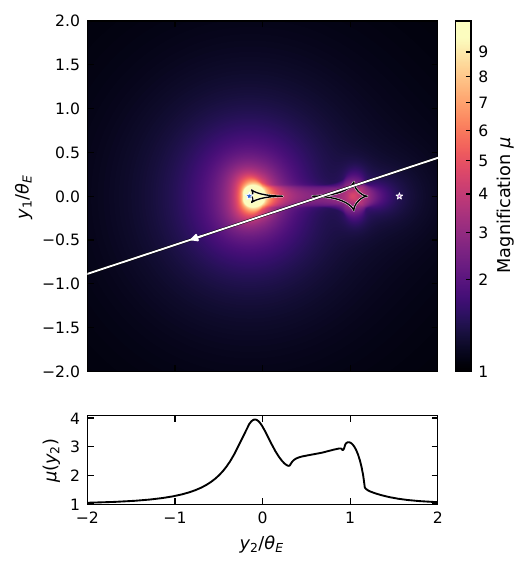}
\caption{Magnification pattern and source trajectory for the adopted binary-lens solution. \textbf{Top}: magnification map with caustics (black curves) and the fitted source trajectory (white line); the arrow indicates the direction of source motion. The color scale gives the magnification $\mu$. \textbf{Bottom}: finite-source magnification along the same trajectory, plotted against the source coordinate $y_2/\theta_{\rm E}$. Source-plane angular coordinates are normalized by the Einstein radius $\theta_{\rm E}$ of the total binary mass.}
\label{fig:geometry}
\end{figure}

The adopted binary-lens model reproduces the two maxima and the intervening minimum in the observed light curve (Figure~\ref{fig:fit}). Along the fitted trajectory, the source encounters two regions of enhanced magnification, with finite-source averaging smoothing these features into broad peaks (Figure~\ref{fig:geometry}). The adopted model reproduces the main double-peaked structure, with $D=-2\ln\mathcal{L}=1364.921$ and $\chi^2_{\rm GP}=337.926$ over the full fitted interval. Table~\ref{tab:lensparams} lists the adopted best-fit values for the seven lens parameters and their 16th--84th percentile intervals (also see Figure~\ref{fig:mcmc_corner}).

\begin{table}[t]
\centering
\caption{Best-fit Binary-lens Parameters\label{tab:lensparams}}
\tabletypesize{\scriptsize}
\setlength{\tabcolsep}{3pt}
\begin{tabular}{lcc}
\hline\hline
Parameter & Symbol & Value (68\% Interval)\\
\hline
Impact parameter & $u_0$ & $0.2134^{+0.0308}_{-0.0205}$\\
Closest approach (MJD) & $t_0$ & $58228.2548^{+0.0143}_{-0.0283}$\\
Einstein time (s) & $t_{\rm E}$ & $28263^{+6232}_{-858}$\\
Normalized source radius & $\rho$ & $0.1205^{+0.0590}_{-0.0314}$\\
Projected separation & $s$ & $1.714^{+0.015}_{-0.208}$\\
Mass ratio & $q$ & $0.0975^{+0.0846}_{-0.0435}$\\
Trajectory angle (rad) & $\alpha$ & $0.3198^{+0.1016}_{-0.0166}$\\
\hline
\end{tabular}
\tablecomments{Intervals are conditional on the adopted local joint GP--lens posterior. The parameters $u_0$, $\rho$, and $s$ are dimensionless.}
\end{table}

\section{Discussion}\label{sec:discussion}

\subsection{Intrinsic Variability and Event Incidence}

The binary-lens interpretation links the two peaks through one lens system and source trajectory while allowing intrinsic evolution of the underlying flare. \citet{Shukla2020} attributed the rapid substructure to magnetic reconnection and plasmoid activity and reported a harder spectrum during the fast flare than during the surrounding envelope. Our time-resolved measurements show hard spectra immediately outside the fast peaks as well, within a broader harder-when-brighter trend. Including fainter, softer phases in the longer envelope intervals can therefore enhance the spectral contrast with the fast flare.

Hydrodynamic simulations provide a possible formation channel for free-floating planetary-mass binaries: tidal bridges produced by encounters between circumstellar disks can fragment into planetary-mass objects, some of which become mutually bound \citep{Fu2025}. Candidate Jupiter-mass binaries have been reported in the Trapezium Cluster \citep{PearsonMcCaughrean2023}, although subsequent spectroscopy classified seven investigated components, each belonging to a different candidate pair, as background sources \citep{Luhman2025JuMBO}. The contribution of planetary-mass binaries to Galactic microlensing depends on their abundance, orbital properties, and dynamical survival in the Galactic field.

For a Galactic lens, the high-latitude sightline to 3C 279 ($b\simeq57^\circ$) rapidly leaves the stellar disk \citep{WangSmith2011}. As an illustrative normalization, an exponential thin disk with a midplane stellar mass density of $0.04\,M_\odot\,{\rm pc}^{-3}$ and a scale height of 300 pc gives $\tau_{\rm thin}\simeq3\times10^{-9}$ toward a distant source along this sightline. Only part of this optical depth can contribute to events near the fitted $t_{\rm E}=0.327$ d: at ordinary Galactic velocities, this timescale selects planetary-mass lenses at typical disk distances or more massive lenses much closer to the observer. The short event rate therefore depends on the abundance of the corresponding binary population.

Let $\tau_{\rm b}$ denote the optical depth defined by one total-mass Einstein radius around each binary center of mass. For a population with a narrow timescale distribution near the fitted value, the geometric encounter rate is \citep{Mroz2020}
\begin{equation}
\Gamma_{\rm b}\simeq 2\tau_{\rm b}/({\pi t_{\rm E}})
\simeq7.1\times10^2\tau_{\rm b}~{\rm yr}^{-1}
\label{eq:discussion-rate}
\end{equation}
per source. The rate of recognizable double-peaked events additionally depends on the trajectory cross-section and the recovery efficiency. The corresponding Galactic mass--distance and source-size relations are illustrated in Appendix~\ref{app:scalings}. The velocities used there specify individual configurations; their relative contributions to the event rate depend on the lens population and detection efficiency.

A foreground lens could instead belong to the galaxy associated with the $z=0.395$ Lyman-limit and metal-line absorber. The sightline passes about $20h^{-1}$ kpc above the disk of an edge-on spiral \citep{Stocke1998}, where $h=H_0/(100~{\rm km\,s}^{-1}\,{\rm Mpc}^{-1})$. For this lens plane, $\Sigma_{\rm crit}=c^2D_{\rm S}/(4\pi G D_{\rm L}D_{\rm LS}) \simeq6.4\times10^3\,M_\odot\,{\rm pc}^{-2}$. Writing $\Sigma_{\rm b}$ for the surface mass density in binaries whose masses and relative angular motions produce events near the observed timescale gives $\tau_{\rm b,fg}=\Sigma_{\rm b}/\Sigma_{\rm crit}$. Equation~\ref{eq:discussion-rate} then yields $\Gamma_{\rm b,fg}\simeq 0.110[\Sigma_{\rm b}/(M_\odot\,{\rm pc}^{-2})]~{\rm yr}^{-1}$. The absorber establishes a possible lens environment, although $\Sigma_{\rm b}$ along this sightline remains unconstrained.

For this foreground population, the expected number of detectable events toward 3C 279 can be written as
\begin{equation}
\begin{split}
\lambda_{279}
&=\Gamma_{\rm b,fg}C_{\rm 2peak}T_{\rm eff}\\
&\simeq2.0
\left(\frac{\Sigma_{\rm b}}{M_\odot\,{\rm pc}^{-2}}\right)
C_{\rm 2peak}
\left(\frac{T_{\rm eff}}{18~{\rm yr}}\right).
\end{split}
\label{eq:discussion-detected-rate}
\end{equation}
Here $C_{\rm 2peak}$ is the effective trajectory-width ratio for producing the double-peaked morphology, relative to the $2R_{\rm E}$ width used in the geometric encounter rate. The corresponding recovery-weighted monitoring time, $T_{\rm eff}=\int\epsilon(t)\,dt$, includes the availability of a sufficiently compact, bright emitting component and the LAT detection efficiency. An expectation of one event requires $\Sigma_{\rm b}C_{\rm 2peak}(T_{\rm eff}/18~{\rm yr}) \simeq0.50\,M_\odot\,{\rm pc}^{-2}$. This is a conditional requirement on the foreground population and effective exposure, rather than a measured surface density.

For a uniformly searched source sample, the total expectation is $\lambda_{\rm surv}=\sum_i\lambda_i$, with a Poisson probability $P(N=1)=\lambda_{\rm surv}\exp(-\lambda_{\rm surv})$ for one detection. The comparison flares analyzed here do not constitute a uniform survey with a calibrated recovery efficiency. A survey-wide interpretation of the presently identified candidate therefore remains conditional on the effective exposure and lens populations along the searched sightlines.

\subsection{Possible Nature of the Lens}
\label{sec:physical}

For angular-diameter distances $D_{\rm L}$, $D_{\rm S}$, and
$D_{\rm LS}$, the Einstein angle of the total binary mass satisfies
\begin{equation}
\theta_{\rm E}^2=\frac{4GM_{\rm tot}}{c^2}
\frac{D_{\rm LS}}{D_{\rm L}D_{\rm S}},
\qquad \theta_{\rm E}=\mu_{\rm rel}t_{\rm E}.
\label{eq:einstein}
\end{equation}
Here $\mu_{\rm rel}$ is the lens--source relative proper motion in the observer frame. The fit constrains the mass ratio $q=M_2/M_1$ and the Einstein time, but not the absolute mass. We adopt $D_{\rm S}=1.31$ Gpc for $z=0.536$ in a flat cosmology with $H_0=70$ km s$^{-1}$ Mpc$^{-1}$ and $\Omega_{\rm m}=0.3$.

In the Galactic limit, $D_{\rm S}\gg D_{\rm L}$,
we write the effective relative transverse speed as
$v_\perp=D_{\rm L}\mu_{\rm rel}$.
Equation~\ref{eq:einstein} becomes
\begin{equation}
\begin{split}
M_{\rm tot}\simeq{}&4.4\times10^{-5}M_\odot
\left(\frac{v_\perp}{100~{\rm km~s}^{-1}}\right)^2\\
&\times\left(\frac{D_{\rm L}}{1~{\rm kpc}}\right)^{-1}
\left(\frac{t_{\rm E}}{2.826\times10^4~{\rm s}}\right)^2.
\end{split}
\label{eq:massscale}
\end{equation}
At $D_{\rm L}=1$ kpc and $v_\perp=100$ km s$^{-1}$, this corresponds to approximately $15M_\oplus$, divided into components near $13M_\oplus$ and $1.3M_\oplus$. The projected separation of the lens components is $a_\perp=sD_{\rm L}\theta_{\rm E}=sv_\perp t_{\rm E}\simeq0.032$~au for this configuration. This is the separation projected onto the sky, rather than the orbital semimajor axis. This illustrative configuration consists of two planetary-mass objects.

For an intervening lens, the relative proper motion can be dominated by the apparent motion of a compact emitting feature in the jet. For a lens at the absorber redshift $z_{\rm L}=0.395$, the adopted distances are $D_{\rm L}=1.10$ Gpc and $D_{\rm LS}=0.307$ Gpc, giving
\begin{equation}
M_{\rm tot}\simeq1.85\times10^{-4}M_\odot
\left(\frac{\mu_{\rm rel}}{20~\mu{\rm as~yr}^{-1}}\right)^2.
\label{eq:discussion-foreground-mass}
\end{equation}
At $\mu_{\rm rel}=0.4$ mas yr$^{-1}$, the mass is about $0.074M_\odot$, with a primary near the stellar/substellar boundary and a companion of about $6.9M_{\rm J}$; the projected separation is about 680 au. A median proper motion of $0.40\pm0.03$~mas~yr$^{-1}$ has been reported for radio knots in 3C~279 \citep{Blinov2021}. The example above assumes a comparable apparent motion for the GeV-emitting region. Its proper motion during the 2018 flare has not been measured. These examples illustrate how the fitted light curve can correspond to lens systems with substantially different component masses and projected separations, depending on the lens distance and relative proper motion.

For a lens in the host galaxy, $D_{\rm L}\simeq D_{\rm S}$, and effective relative transverse speeds of 100--1000 km s$^{-1}$ in the host frame would require an extremely compact emitting region, with $R_{\gamma,\perp}\simeq\rho v_{\rm eff}t_{\rm E}/(1+z)\simeq2.2\times10^{10}$--$2.2\times10^{11}$ cm. Such a small size would make it difficult for gamma-ray photons to escape, disfavoring the host-lens interpretation.

\subsection{Size of the Gamma-Ray Emitting Region}

The fitted finite-source parameter corresponds to an observer-frame source-radius crossing time of $t_\star=\rho t_{\rm E}\simeq3.41\times10^3$~s, or 0.95~hr. The observed peak widths also depend on the trajectory orientation and the local caustic structure. The projected physical radius of the magnified component is
\begin{equation}
R_{\gamma,\perp}=D_{\rm S}\rho\mu_{\rm rel}t_{\rm E}
\simeq2.8~{\rm au}
\left(\frac{\mu_{\rm rel}}{20~\mu{\rm as~yr}^{-1}}\right).
\label{eq:sourcesize}
\end{equation}
No further redshift factor enters because $\mu_{\rm rel}$ is defined in observer time. The intervening-lens example at 0.4 mas yr$^{-1}$ gives about 57 au. The 1-kpc Galactic example instead has $\mu_{\rm rel}\simeq21$ mas yr$^{-1}$ and $R_{\gamma,\perp}\simeq3000$ au ($4.5\times10^{16}$ cm). These radii describe the projected magnified component, not its distance along the jet or the full jet cross-section. Microlensing of B0218+357 has yielded gamma-ray source-size constraints in a different lens geometry \citep{VovkNeronov2016}.

For the adopted $t_{\rm E}$, $\rho$, and $D_{\rm S}$, eliminating $D_{\rm L}$ in the Galactic limit gives
\begin{equation}
M_{\rm tot}\simeq4.4\times10^{-5}M_\odot
\left(\frac{v_\perp}{100~{\rm km~s}^{-1}}\right)
\left(\frac{R_{\gamma,\perp}}{3000~{\rm au}}\right).
\label{eq:discussion-mass-radius}
\end{equation}
Consequently, increasing the lens mass at a fixed speed also increases the inferred source radius. A $0.02M_\odot$ binary at 100 km s$^{-1}$ would lie at about 2.2 pc and imply a radius of 6.6 pc, much larger than in the planetary-mass example.

Under the lensing interpretation, the sharp peaks need not trace equally rapid intrinsic variability. If the magnified component also produces the broad intrinsic envelope, its size can be compared with the causality limit inferred from that envelope. Approximating this component as a coherently varying sphere in its comoving frame, we identify $R_{\gamma,\perp}\simeq R'$ and obtain
\begin{equation}
\begin{split}
R'&\lesssim\frac{c\delta t_{\rm int}}{1+z}\\
&\simeq5.2\times10^{16}
\left(\frac{\delta}{20}\right)
\left(\frac{t_{\rm int}}{1.34\times10^5~{\rm s}}\right)
{\rm cm},
\end{split}
\label{eq:discussion-causality}
\end{equation}
where $R'$ is the comoving radius and $\delta$ is the Doppler factor. We use the fitted rise parameter of the smooth flare component, $\tau_{\rm rise}=1.34\times10^5$~s, as the characteristic observer-frame envelope timescale $t_{\rm int}$.
At $\delta\simeq20$, the 1-kpc planetary-mass configuration lies near this causality limit, whereas the nearby $0.02M_\odot$ configuration exceeds it by approximately a factor of 390. Within this single-region approximation, the intrinsic rise timescale therefore places an additional constraint on the Galactic lens mass at a specified transverse speed.

Finally, $\rho$ describes a band-integrated source. Gravitational deflection is wavelength independent, but an energy-dependent source size can produce different caustic smoothing and magnification across gamma-ray bands \citep{Torres2003}. Similar photon indices are therefore compatible with lensing without establishing achromaticity. The finite-source fit thus constrains the effective radius of the adopted uniform-brightness disk in units of $\theta_{\rm E}$, while its physical scale remains dependent on the lens distance and relative proper motion.

\section{Conclusions}\label{sec:summary}

A finite-source binary-lens model jointly fitted with intrinsic Gaussian-process variability reproduces the principal double-peaked structure of the April 2018 gamma-ray flare of 3C~279, with $q\simeq0.0975$ and $t_{\rm E}\simeq0.327$~d. The inferred lens masses and source radius depend on the lens distance and relative proper motion, allowing distinct Galactic and intervening-lens configurations. Binary microlensing remains a testable interpretation of this event, rather than a uniquely established origin of the sharp peaks. Under this interpretation, the finite-source fit constrains the size of the magnified GeV-emitting component relative to the Einstein radius.


\clearpage
\appendix
\restartappendixnumbering

\section{LAT Analysis, Temporal and Spectral Statistics}
\label{app:lat}
\resetapptablenumbers

We analyzed \textit{Fermi}-LAT SOURCE-class events within a $12^\circ$ region of interest centered on 3C 279, using the P8R3\_SOURCE\_V3 instrument response functions and excluding events with zenith angles above $90^\circ$. The region model included sources from 4FGL-DR3 \citep{Abdollahi2022} and the corresponding Galactic and isotropic diffuse components. After optimizing the model over the full interval, we refitted the normalization and spectral shape of 3C 279 in each shorter time bin, while holding the parameters of the other sources fixed at their full-interval values. Each flux estimate was derived from a likelihood fit to the LAT events within the respective bin.

Adaptive binning followed \citet{Lott2012}. We also used fixed 1000, 2500, and 6000 s bins to examine the temporal structure of the flare. The light-curve fit is based on 444 retained 2500 s flux estimates with $\mathrm{TS}>4$, after excluding two estimates that failed to meet this threshold.

We also examined the available contemporaneous X-ray observations \citep{StrohFalcone2013,Krimm2013} for counterparts to the sharp GeV peaks. The sparse Swift/XRT snapshots lack the cadence needed to resolve corresponding rapid variability, while the low significance of the Swift/BAT measurements limits the search for a hard-X-ray counterpart. These observations therefore do not establish whether the sharp GeV peaks are accompanied by similar X-ray features.

Under the microlensing interpretation, the X-ray counterpart could be weaker and smoother if the X-ray emission extends over a larger region than the magnified GeV component, owing to finite-source averaging \citep{Torres2003}. Emission from additional components that are weakly affected by the lens could further dilute the observed variability. The expected X-ray response thus depends on both the spatial distribution of the emission and the contribution of the compact magnified component to the total X-ray flux.

\subsection{Comparison Flare Sample}
\label{app:comparison}

Table~\ref{tab:comparison-flares} lists 58 selected flare fits: 54 from other blazars, three other flares of 3C 279, and the 2018 event (ID 58). The MJD ranges identify the selected flare intervals. Each flare is fitted with the smooth double-exponential profile of Equation~\ref{eq:flare}. The tabulated amplitude is $A=\max(F_{\rm fit})/\langle F_{\rm obs}\rangle$, and the empirical flare score is $S=A/(\tau_{\rm rise}+\tau_{\rm decay})$, with the timescales expressed in days. For the 2018 event, $A=3.966$, $\tau_{\rm rise}=0.372$ d, and $\tau_{\rm decay}=0.307$ d give $S=5.84$~d$^{-1}$.

\begin{figure}[ht]
\centering
\includegraphics[width=0.9\linewidth]{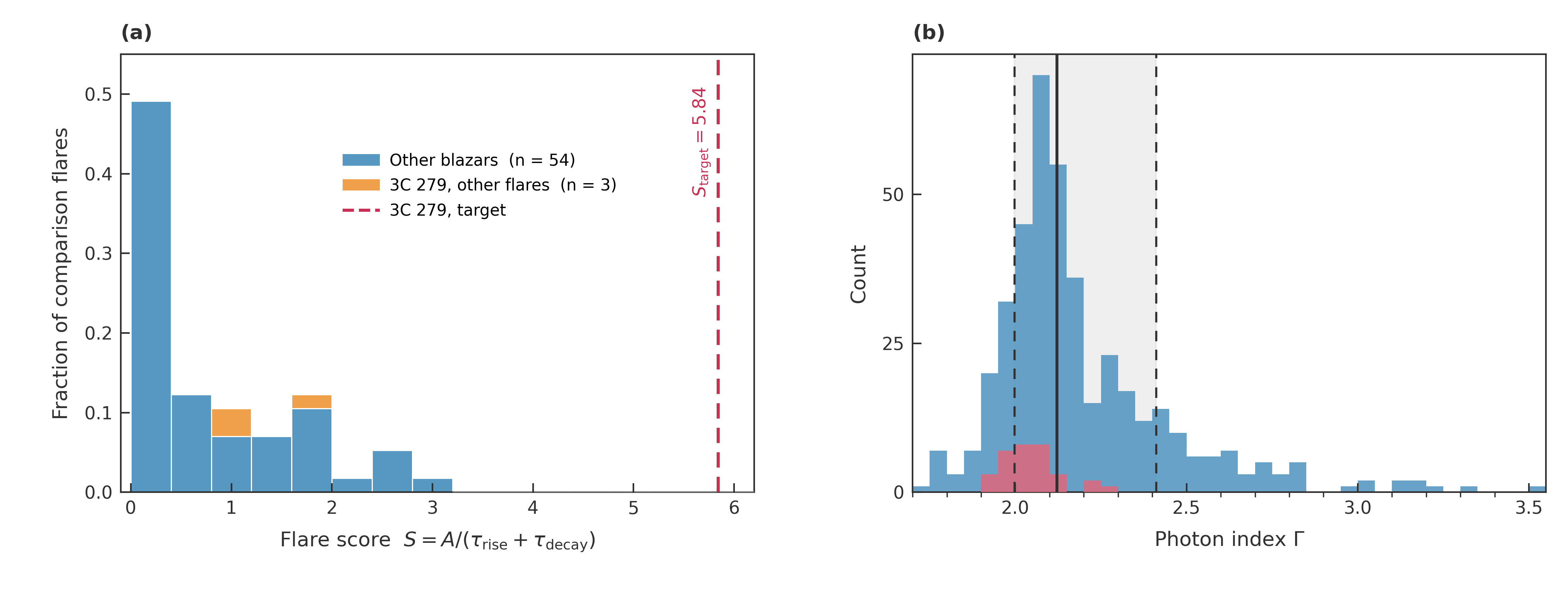}
\caption{\textbf{(a)} Empirical flare-score distribution for the 2018 3C~279 event (red dashed line), three other 3C~279 flares (orange), and 54 flares from other blazars (blue; Table~\ref{tab:comparison-flares}). \textbf{(b)} Photon-index distributions over MJD 58219.60--58232.47. Red and blue show 32 estimates during the putative lensing interval and 412 outside it, respectively, overlaid without stacking. The black solid and dashed lines indicate the median and 16th--84th percentiles of the blue sample, with the percentile interval shaded.}
\label{fig:flare-score-index}
\end{figure}

Figure~\ref{fig:flare-score-index}(a) compares the scores of the 2018 event with those of the other 3C 279 flares and the other-blazar sample. The 2018 event has the highest score in this sample, whereas most comparison flares have $S<3$~d$^{-1}$, indicating a more temporally concentrated profile for the target event. The sample comprises individually selected fitting windows and does not constitute a complete, uniformly selected flare census.

\begin{deluxetable*}{r l r r r r@{\hspace{0.45cm}}r l r r r r}
\tabletypesize{\scriptsize}
\tablecaption{Flare Comparison Sample\label{tab:comparison-flares}}
\tablehead{
\colhead{ID} & \colhead{Source} & \colhead{MJD} &
\colhead{$A$} & \colhead{$\tau_{\rm rise}$} &
\colhead{$\tau_{\rm decay}$} &
\colhead{ID} & \colhead{Source} & \colhead{MJD} &
\colhead{$A$} & \colhead{$\tau_{\rm rise}$} &
\colhead{$\tau_{\rm decay}$} \\
\colhead{} & \colhead{} & \colhead{} &
\colhead{} & \colhead{(d)} & \colhead{(d)} &
\colhead{} & \colhead{} & \colhead{} &
\colhead{} & \colhead{(d)} & \colhead{(d)}
}
\startdata
1 & 0108.6+0134 & 57575--57592 & 2.105 & 1.043 & 3.709 & 30 & 1512.8-0906 & 56561--56573 & 2.631 & 0.091 & 1.248 \\
2 & 0108.6+0134 & 59673--59716 & 2.485 & 19.233 & 18.126 & 31 & 1512.8-0906 & 57112--57119 & 2.773 & 1.363 & 2.072 \\
3 & 0112.8+3208 & 60074--60127 & 3.016 & 1.844 & 12.600 & 32 & 1512.8-0906 & 57239--57254 & 3.640 & 1.076 & 3.572 \\
4 & 0205.0-1700 & 59696--59704 & 8.638 & 1.312 & 1.549 & 33 & 1512.8-0906 & 57633--57636 & 2.334 & 0.600 & 0.312 \\
5 & 0237.8+2848 & 55874--55905 & 1.915 & 2.449 & 14.948 & 34 & 1635.2+3808 & 56484--56527 & 2.209 & 8.793 & 17.916 \\
6 & 0237.8+2848 & 58285--58327 & 2.127 & 7.653 & 1.823 & 35 & 1642.9+3948 & 59829--59837 & 5.705 & 1.320 & 1.492 \\
7 & 0403.9-3605 & 60004--60010 & 1.667 & 2.422 & 2.134 & 36 & 1833.6-2103 & 55483--55489 & 2.756 & 0.790 & 16.471 \\
8 & 0505.3+0459 & 57092--57113 & 3.322 & 3.436 & 3.881 & 37 & 1833.6-2103 & 58542--58634 & 3.144 & 20.654 & 18.854 \\
9 & 0739.2+0137 & 58628--58641 & 1.308 & 5.370 & 1.728 & 38 & 1833.6-2103 & 59892--59918 & 2.011 & 8.801 & 8.037 \\
10 & 0904.9-5734 & 58941--58947 & 5.087 & 0.601 & 1.293 & 39 & 2025.6-0735 & 57483--57500 & 3.819 & 5.081 & 1.658 \\
11 & 0904.9-5734 & 59424--59429 & 3.651 & 0.222 & 1.731 & 40 & 2201.8+5048 & 56108--56121 & 4.963 & 0.092 & 2.433 \\
12 & 0904.9-5734 & 59584--59596 & 1.615 & 0.852 & 3.821 & 41 & 2202.7+4216 & 59115--59126 & 1.565 & 0.797 & 0.616 \\
13 & 0904.9-5734 & 59717--59730 & 2.367 & 0.238 & 12.756 & 42 & 2202.7+4216 & 59419--59441 & 2.777 & 9.222 & 6.619 \\
14 & 0904.9-5734 & 59847--59856 & 2.917 & 0.561 & 2.055 & 43 & 2202.7+4216 & 59637--59659 & 1.728 & 11.292 & 5.545 \\
15 & 1159.5+2914 & 58095--58111 & 1.345 & 7.435 & 0.964 & 44 & 2232.6+1143 & 57419--57461 & 3.485 & 6.829 & 1.482 \\
16 & 1159.5+2914 & 59476--59487 & 2.143 & 0.979 & 5.059 & 45 & 2232.6+1143 & 57702--57777 & 4.353 & 15.812 & 6.203 \\
17 & 1159.5+2914 & 59942--59967 & 3.188 & 4.273 & 4.581 & 46 & 2232.6+1143 & 58137--58155 & 1.845 & 1.890 & 2.372 \\
18 & 1224.9+2122 & 55353--55402 & 2.783 & 2.829 & 20.072 & 47 & 2253.9+1609 & 54712--54737 & 1.597 & 0.906 & 11.107 \\
19 & 1224.9+2122 & 56695--56709 & 2.421 & 3.565 & 2.534 & 48 & 2253.9+1609 & 55276--55317 & 2.514 & 7.120 & 10.849 \\
20 & 1229.0+0202 & 54738--54753 & 1.949 & 6.281 & 3.959 & 49 & 2253.9+1609 & 55514--55526 & 6.988 & 0.808 & 2.700 \\
21 & 1332.0-0509 & 55499--55510 & 3.067 & 3.819 & 0.227 & 50 & 2253.9+1609 & 56827--56838 & 3.313 & 0.620 & 2.031 \\
22 & 1427.9-4206 & 56362--56453 & 1.364 & 43.797 & 22.086 & 51 & 2253.9+1609 & 57067--57073 & 1.951 & 0.551 & 1.709 \\
23 & 1427.9-4206 & 59913--59953 & 2.379 & 6.896 & 13.519 & 52 & 2253.9+1609 & 57319--57337 & 2.501 & 1.069 & 0.285 \\
24 & 1504.4+1029 & 54781--54824 & 1.756 & 5.956 & 12.196 & 53 & 2253.9+1609 & 57404--57411 & 2.234 & 0.375 & 1.079 \\
25 & 1504.4+1029 & 57195--57239 & 2.050 & 7.763 & 13.605 & 54 & 2253.9+1609 & 57595--57600 & 2.723 & 0.126 & 0.882 \\
26 & 1512.8-0906 & 54913--54919 & 2.628 & 1.447 & 0.552 & 55 & 3C~279 & 57186--57191 & 6.990 & 2.771 & 1.301 \\
27 & 1512.8-0906 & 55201--55213 & 2.310 & 1.133 & 0.366 & 56 & 3C~279 & 58128--58145 & 3.997 & 1.153 & 2.352 \\
28 & 1512.8-0906 & 55865--55872 & 3.546 & 0.611 & 1.348 & 57 & 3C~279 & 58262--58279 & 3.835 & 1.283 & 2.102 \\
29 & 1512.8-0906 & 55996--56008 & 1.529 & 3.361 & 1.106 & 58 & 3C~279 & 58227--58229 & 3.966 & 0.372 & 0.307 \\
\enddata
\tablecomments{$A=\max(F_{\rm fit})/\langle F_{\rm obs}\rangle$ is the dimensionless normalized peak amplitude. The listed MJD ranges identify flare intervals. Entries 1--54 are other-blazar flares, 55--57 are other 3C 279 flares, and 58 is the target.}
\end{deluxetable*}

\onecolumngrid

\subsection{Photon-index Distribution}
\label{app:spectral-index}

Figure~\ref{fig:flare-score-index}(b) compares the 32 photon-index estimates during the sharp peaks, MJD 58227.67319--58228.62213, with the 412 estimates outside this interval within the broader MJD 58219.60--58232.47 window. The estimates during the sharp peaks are concentrated in the same central range as those from the rest of the flare, complementing the time-resolved spectral panel in Figure~\ref{fig:fit}. The histogram summarizes the fitted index values without incorporating their individual uncertainties and provides a descriptive comparison of the two samples.

\citet{Shukla2020} also discussed the spectral index behavior during this period. Notably, their analysis utilized a global log-parabola model to calculate the flux in each time bin, yielding $\alpha$ as the spectral index descriptor. In contrast, given the limited photon statistics within individual bins, our analysis adopts a power-law model, such that we compare the photon index $\Gamma$ instead. To ensure consistency, we re-evaluated the spectrum of the ``fast flare'' (MJD 58227.76--58228.40) from \citet{Shukla2020} using a log-parabola model, obtaining $\alpha = 1.8675 \pm 0.0256$, which is in excellent agreement with their reported value. Furthermore, we suggest that the significant difference between the spectral index during the fast flare and that of the broader envelope claimed in their work is primarily an averaging effect driven by the much softer spectra during the low-flux phases on either side of the envelope, a point we also elaborate on in the main text.

%
%

\onecolumngrid

\section{Gaussian-process Analysis and Binary-lens Fitting}
\label{app:diagnostics}
\resetapptablenumbers

\subsection{Gaussian-process Fit to the Observed Light Curve}
\label{app:gp}

Gaussian processes (GPs) utilizing a stochastically driven, damped simple-harmonic-oscillator (SHO) kernel provide a robust framework for modeling astronomical time series \citep{ForemanMackey2017}. Such GP models have been increasingly employed to characterize the long-term gamma-ray variability of jetted AGNs \citep{Zhang2022GammaGP}. Furthermore, applications to individual bright flares of sources like 3C 454.3 and 3C 279 have demonstrated that the SHO kernel can successfully capture variability features beyond the simpler damped-random-walk prescription \citep{Zhang2025FlareGP}. Following this approach, we fitted the observed light curve directly with a GP using an SHO kernel. In this fit, the time-dependent emission was described by the GP, without a prescribed double-exponential flare profile or a lensing magnification term.

\begin{figure*}[ht]
\centering
\begin{minipage}[t]{0.48\textwidth}
\vspace{0pt}
\centering
\textbf{(a)}\par\smallskip
\includegraphics[width=\linewidth]{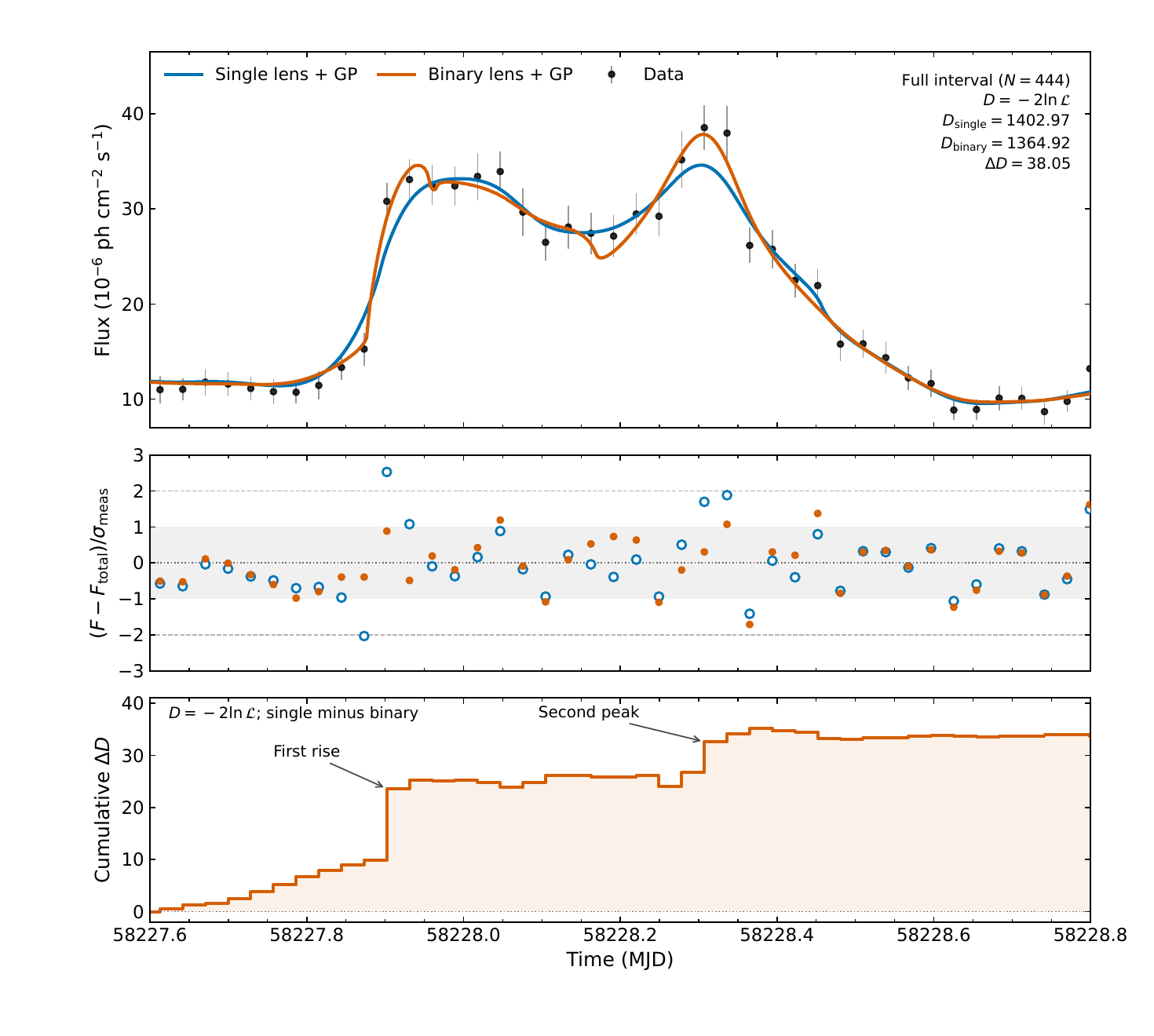}
\end{minipage}\hfill
\begin{minipage}[t]{0.50\textwidth}
\vspace{0pt}
\centering
\textbf{(b)}\par\smallskip
\includegraphics[width=\linewidth]{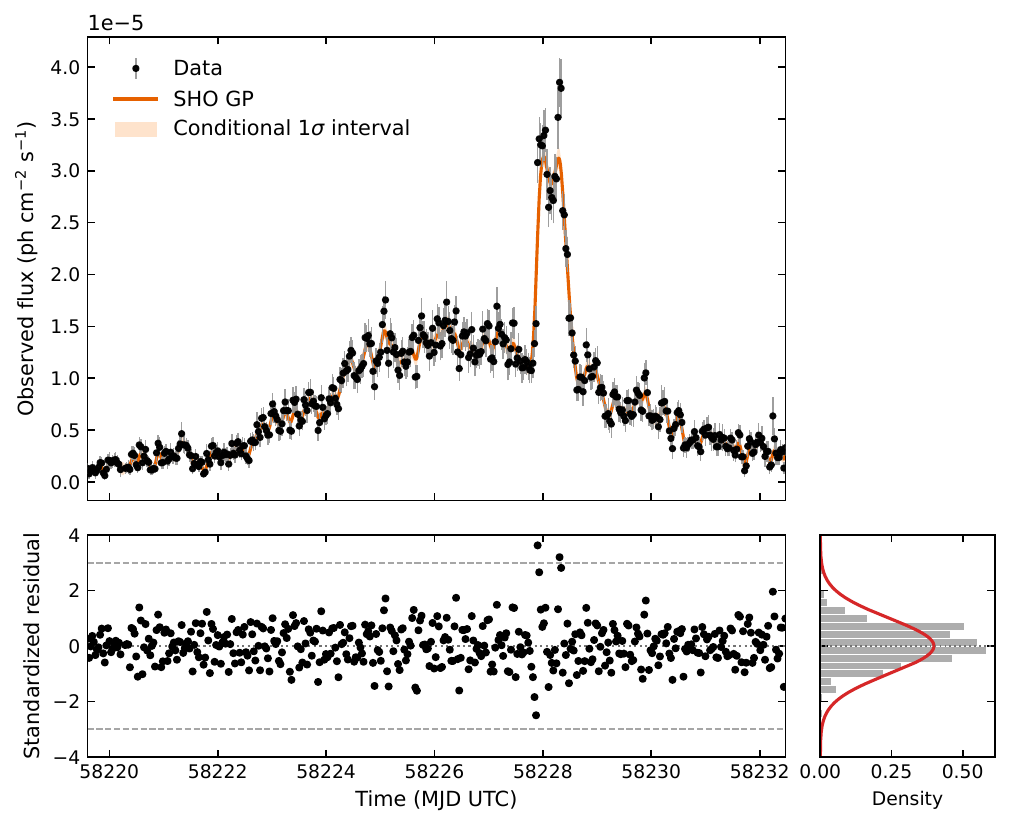}
\end{minipage}
\caption{Lens-model comparison and a GP-only description of the LAT flare.
\textbf{(a)} Single-lens+GP (blue) and binary-lens+GP (orange) fits around the sharp peaks. The upper panel shows the measured fluxes with uncertainties (black points) and the total model predictions, including the GP contribution. The middle panel shows residuals divided by the measurement uncertainties, $\sigma_{\rm meas}$; open blue and filled orange circles correspond to the single- and binary-lens models, respectively. The shaded band marks $\pm1$, and the dashed lines mark $\pm2$. The lower panel shows the cumulative deviance difference, $\Delta D=D_{\rm single}-D_{\rm binary}$, over the displayed interval, where $D=-2\ln\mathcal{L}$; positive values favor the binary-lens model. The arrows mark the first rapid rise and the second peak. The upper-right annotation reports statistics for all 444 flux measurements. \textbf{(b)} GP-only fit over the full flare interval using an SHO kernel. The upper panel shows the observed fluxes (black points), conditional GP mean (orange), and conditional $1\sigma$ interval (shading). The lower panels show the standardized residuals and their distribution, with a Gaussian reference curve in red. The two panels use different time ranges and flux-axis scalings.}
\label{fig:gp_diagnostic}
\end{figure*}

Figure~\ref{fig:gp_diagnostic}(b) shows the conditional GP mean and its $1\sigma$ interval, together with the standardized residuals and their distribution. The GP follows the broad evolution of the flare and much of the rapid brightening, but underpredicts several of the highest flux measurements. The largest positive residuals occur near the sharp peaks. These localized deviations indicate that the sharpest structure is not fully captured by the fitted SHO process.


We also compare single- and binary-lens models, each supplemented by a GP, using the same 444 flux measurements (Figure~\ref{fig:gp_diagnostic}(a)). The fits are evaluated using the GP marginal likelihood, with the latent process integrated out. The binary-lens model gives $D=-2\ln\mathcal{L}=1364.92$, compared with $D=1402.97$ for the single-lens model, an improvement of $\Delta D=38.05$. The binary-lens model better reproduces the steep rise of the first peak and the height of the second peak, while the predictions are similar over much of the remaining interval.

Counting the fitted mean-model parameters and GP hyperparameters, the binary-lens model has three additional free parameters. With differences defined as single minus binary, the conventional information criteria give $\Delta{\rm AIC}=\Delta D-2\Delta k=32.05$ and $\Delta{\rm BIC}=\Delta D-\Delta k\ln N=19.76$. These approximate criteria favor the binary-lens model.

\subsection{Parameter Search and Local Posterior}
\label{app:lenssearch}

\begin{figure}[htbp]
\centering
\includegraphics[width=0.6\textwidth]{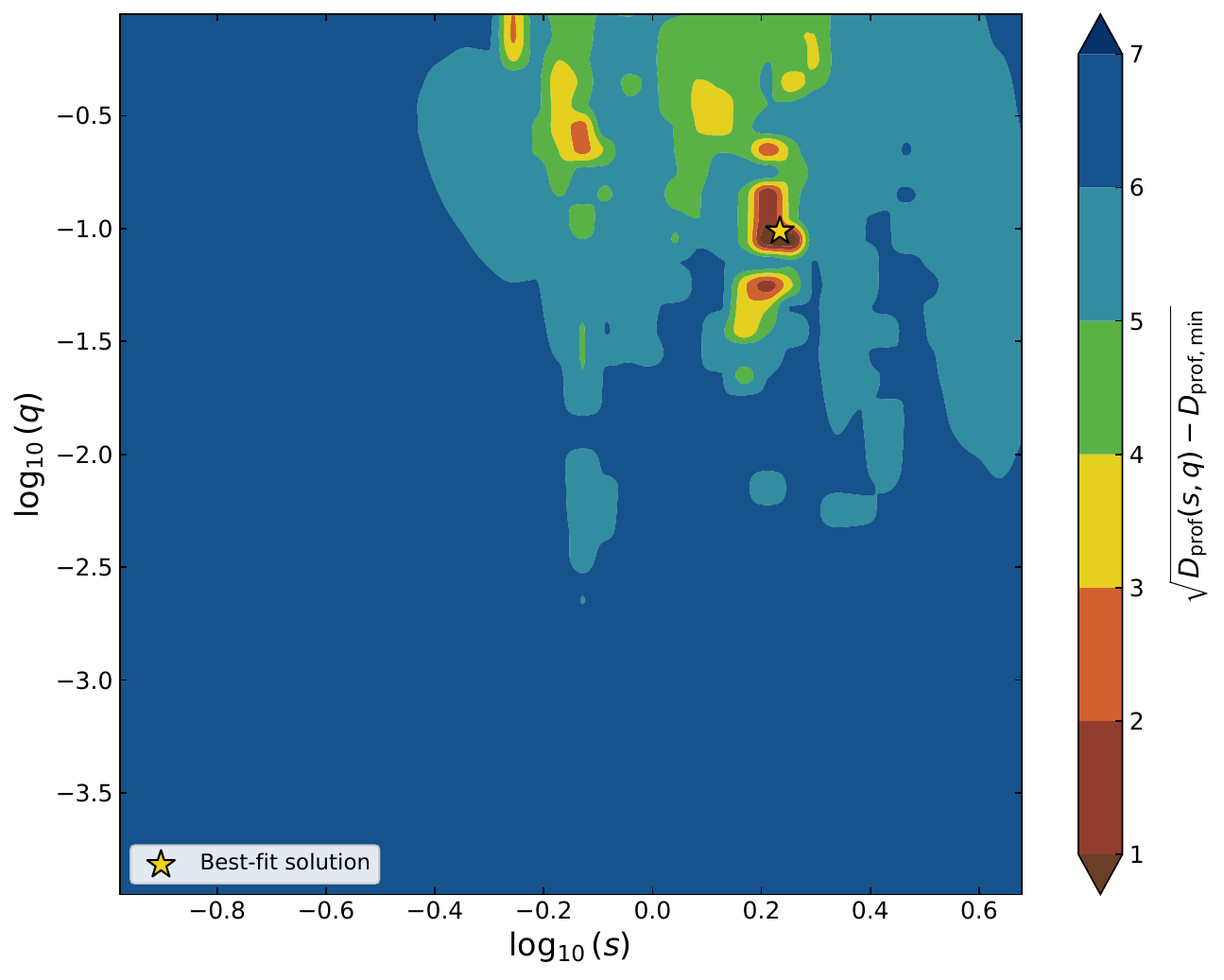}
\caption{Profiled fixed-$(s,q)$ surface, with the adopted best-fit solution marked by a star. The color scale is $[D_{\rm prof}(s,q)-D_{\rm prof,min}]^{1/2}$, where $D=-2\ln\mathcal{L}$, not a calibrated Gaussian significance.}
\label{fig:grid_profile}
\end{figure}

We searched a $40\times40$ grid spanning $-1\leq\log_{10}s\leq0.7$ and $-4\leq\log_{10}q\leq0$, with 16 initial trajectory angles over $[0,2\pi)$ and three initial Einstein timescales of $2.5\times10^4$, $1.0\times10^5$, and $3.5\times10^5$~s at each grid point. For each fixed $(s,q)$ pair, the remaining lens, flare, and GP parameters were optimized using bounded L-BFGS-B followed by CMA-ES \citep{Hansen2001}. We constrained the flare parameters to remain near those describing the broad intrinsic envelope. At each grid point, we retained the lowest $D=-2\ln\mathcal{L}$ found after optimization over the remaining parameters (Figure~\ref{fig:grid_profile}).

The 3000 configurations with the lowest $D$ were further optimized, and representative solutions were subsequently explored through local MCMC sampling with \texttt{emcee} \citep{ForemanMackey2013}. We used 48 walkers per ensemble, discarded the first 1000 steps as burn-in, and extended the production chains in blocks of 500 steps. The parameter bounds were unchanged from the search; $\rho$, $s$, and $q$ were sampled in logarithmic coordinates, $\alpha$ over a full $2\pi$ interval, and the SHO hyperparameters in natural-log coordinates.

\begin{figure*}[htbp]
\centering
\includegraphics[width=0.88\textwidth]{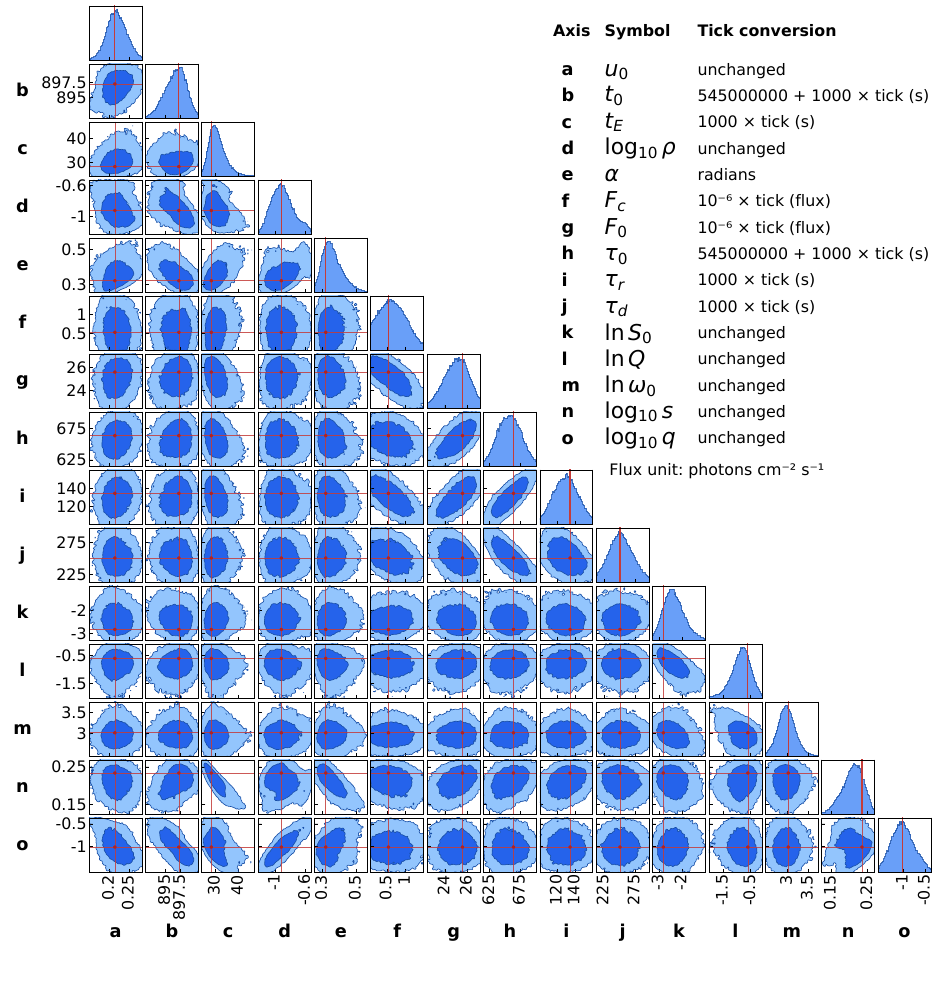}
\caption{Local posterior distributions of the 15 lens, intrinsic-flare, and SHO parameters. Shaded contours show the 68\% and 95\% two-dimensional density regions, and red lines mark the adopted best-fit parameters used in Figures~\ref{fig:fit} and \ref{fig:geometry}. The sampled coordinates include $\log_{10}\rho$, $\log_{10}s$, and $\log_{10}q$; the parameter table reports their transformed values.}
\label{fig:mcmc_corner}
\end{figure*}

Figure~\ref{fig:mcmc_corner} shows the joint and marginal posterior distributions for the adopted best-fit solution found through the grid search and subsequent MCMC analysis. The best-fit lens model gives $D=1364.921$ and $\chi^2_{\rm GP}=337.926$. The quoted parameter intervals are the 16th--84th percentiles of the adopted local posterior mode.

\section{Galactic Lens and Source Scalings}
\label{app:scalings}
\resetapptablenumbers

For a Galactic lens, $D_{\rm L}\ll D_{\rm S}$ and $D_{\rm LS}\simeq D_{\rm S}$. Here $D_{\rm S}$ is the angular-diameter distance to 3C~279, and $v_\perp=D_{\rm L}\mu_{\rm rel}$ is the effective relative transverse speed projected onto the lens plane. In this limit, Equations~\ref{eq:massscale} and \ref{eq:sourcesize} give
\begin{equation}
D_{\rm L}\simeq c^2v_\perp^2t_{\rm E}^2/(4GM_{\rm tot})
\simeq 43.8~{\rm pc}\,
[v_\perp/(100~{\rm km~s}^{-1})]^2
[M_{\rm tot}/(10^{-3}M_\odot)]^{-1}.
\label{eq:lens-mass-distance-locus}
\end{equation}
for $t_{\rm E}=2.826\times10^4$~s, where $M_{\rm tot}=M_1+M_2$. The corresponding source-plane radius is
\begin{equation}
R_{\gamma,\perp}=\rho D_{\rm S}\theta_{\rm E}
\simeq \rho D_{\rm S}[4GM_{\rm tot}/(c^2D_{\rm L})]^{1/2}
\simeq 4G\rho D_{\rm S}M_{\rm tot}/(c^2v_\perp t_{\rm E}).
\label{eq:lens-source-radius-locus}
\end{equation}
We use $\rho=0.1205$ and $D_{\rm S}=1.31$~Gpc. For example, $M_{\rm tot}=10^{-3}M_\odot$ and $v_\perp=100$~km~s$^{-1}$ give $D_{\rm L}\simeq43.8$~pc and $R_{\gamma,\perp}\simeq1.02\times10^{18}$~cm.

\begin{figure}[htbp]
\centering
\includegraphics[width=0.6\linewidth]{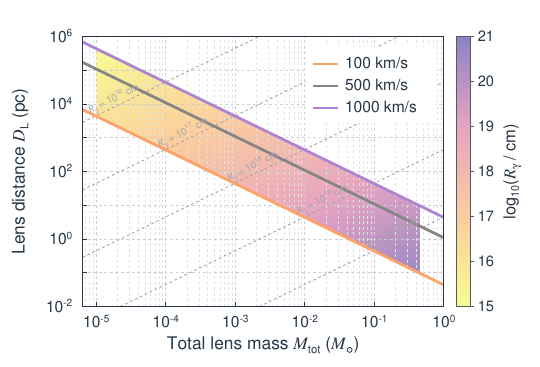}
\caption{Galactic lens mass--distance relations and the corresponding source-plane radius for $t_{\rm E}=2.826\times10^4$~s, $\rho=0.1205$, and $D_{\rm S}=1.31$~Gpc. The orange, gray, and purple curves correspond to effective transverse speeds of 100, 500, and 1000~km~s$^{-1}$, respectively. The color scale shows $\log_{10}(R_{\gamma,\perp}/{\rm cm})$. The colored region spans total lens masses $10^{-5}$--$0.45\,M_\odot$ and transverse speeds between 100 and 1000~km~s$^{-1}$. Dark dashed lines show contours of constant projected source radius across the mass–distance plane.}
\label{fig:lens-mass-source-scalings}
\end{figure}

At fixed $t_{\rm E}$ and $v_\perp$, a larger lens mass corresponds to a smaller lens distance and a larger source-plane radius. Figure~\ref{fig:lens-mass-source-scalings} illustrates these relations for transverse speeds of 100, 500, and 1000~km~s$^{-1}$. The configurations are shown without population weighting; their relative contributions to the event rate depend on the lens density, mass and velocity distributions, and binary properties.

\FloatBarrier
\bibliographystyle{aasjournalv7.1}
\bibliography{references}
\end{document}